\documentclass[11pt,twoside,onecolumn]{article}
\usepackage[]{latexsym}
\usepackage{epsfig}
\usepackage{amsmath,amssymb}
\newcommand{\be}{\begin{eqnarray}}
\newcommand{\ee}{\end{eqnarray}}
\newcommand{\ud}{\mathrm{d}}

\newcommand{\gn}{G_{\rm N}}
\newcommand{\lp}{\ell_{\rm p}}
\newcommand{\mpl}{m_{\rm p}}

\title{\bf Quantum Harmonic Black Holes}
\author{Roberto~Casadio$^{a,b}$\thanks{roberto.casadio@bo.infn.it}
$\ $
and
Alessio Orlandi$^{a,b}$\thanks{alessio.j.orlandi@gmail.com}
\\
\null
\\
$^a${\em Dipartimento di Fisica e Astronomia, Universit\`a di Bologna}
\\
{\em via Irnerio~46, 40126 Bologna, Italy}
\\
\\
$^c${\em Istituto Nazionale di Fisica Nucleare, Sezione di Bologna}
\\
{\em via Irnerio~46, 40126 Bologna, Italy}
}
\begin{document}
\maketitle
\begin{abstract}
Inspired by the recent conjecture that black holes are condensates of gravitons,
we investigate a simple model for the black hole degrees of freedom that is consistent
both from the point of view of Quantum mechanics and of General Relativity.
Since the two perspectives should ``converge'' into a unified picture for small,
Planck size, objects, we expect our construction is a useful step for understanding
the physics of microscopic, quantum black holes.
In particular, we show that a harmonically trapped condensate gives rise to
two horizons, whereas the extremal case (corresponding to a remnant with
vanishing Hawking temperature) naturally falls out of its spectrum.
\end{abstract}
%
%
%\pacs{04.70.Dy,04.70.-s,04.60.-m}
%
%
\section{Introduction and motivation}
\label{secIntro}
\setcounter{equation}{0}
One of the major mysteries in modern theoretical physics is to understand
what are the internal degrees of freedom of black holes.
This issue becomes particularly relevant in any attempt to develop a
quantum theory which incorporates gravity along with the other forces of nature.
Of course, without experimental inputs, our best starting point is the
classical description of black holes provided by General Relativity~\cite{chandra},
along with well established semiclassical results, such as the
predicted Hawking radiation~\cite{hawking}.
\par
It was recently proposed by Dvali and Gomez that black holes are
Bose-Einstein Condensates (BECs) of gravitons at a critical point,
with Bogoliubov modes that become degenerate and nearly gapless
representing the holographic quantum degrees of freedom responsible
for the black hole entropy and the information storage~\cite{gomez}.
In order to support this view, they consider a collection of objects (gravitons)
interacting via Newtonian gravity,
\be
V_{\rm N}
\sim
-\frac{\gn\,\mu}{r}
\ ,
\ee
and whose effective mass $\mu$ is related to their characteristic quantum mechanical
size via the Compton/de~Broglie wavelength,
\be
\ell
\simeq
\frac{\hbar}{\mu}
=
\lp\,\frac{\mpl}{\mu}
\ .
\ee
These bosons can superpose and form a ``ball'' of radius $\ell$,
and total energy $M=N\,\mu$, where $N$ is the total number of constituents.
Within the Newtonian approximation, there is then a value of $N$
for which the whole system becomes a black hole.
In details, given the coupling constant
\be
\alpha
=
\frac{\lp^2}{\ell^2}
=
\frac{\mu^2}{\mpl^2}
\ ,
\ee
there exists an integer $N$ such that no constituent can escape the 
gravitational well it contributed to create, and which can be approximately
described by the potential
\be
U(r)
\simeq
V_{\rm N}(\ell)
\simeq
-N\,\alpha\,\frac{ \hbar}{\ell}\,\Theta(\ell-r)
\ ,
\label{Udvali}
\ee
where $\Theta$ is the Heaviside step function.
This implies that components in the depleting region
are ``marginally bound'', 
\be
E_K + U
\simeq
0
\ ,
\label{energy0}
\ee
where the kinetic energy is given by $E_K\simeq \mu$.
This energy balance yields the ``maximal packing''
\be
N\,\alpha = 1
\ .
\label{maxP}
\ee
Consequently, the effective boson mass and total mass of the
balk hole scale according
to
\be
\mu
\simeq
\frac{\mpl}{\sqrt{N}}
\quad
{\rm and}
\quad
M
=
N\,\mu
\simeq
\sqrt{N}\,\mpl
\ .
\label{Max}
\ee
Note that one has here assumed the ball is of size $\ell$
(since bosons superpose) and, therefore, the constituents will
interact at a maximum distance of order $r\sim\ell$, with fixed $\ell$.
The Hawking radiation and the negative specific heat
spontaneously result from quantum depletion of the condensate
for the states satisfying Eq.~\eqref{energy0}.
This description is partly Quantum Mechanics and partly classical Newtonian physics,
but no General Relativity is involved, in that geometry does not appear in the argument.
\par
In this work, we will show how this picture, which draws from the conjectured
UV-self-complete\-ness of gravity~\cite{dvali}, can be both improved within
Quantum Mechanics and reconciled with the usual geometric description of
space-time in General Relativity.
Some considerations about the possible existence of remnants will also
follow.
We shall use units with $c=1$, $\hbar=\lp\,\mpl$ and the Newton constant
$\gn=\lp/\mpl$.
\section{Quantum mechanical model}
\label{secModel}
\setcounter{equation}{0}
To summarise, Ref.~\cite{gomez} assumes that a black hole is a BEC, trapped
in a gravitational well described by the simple potential~\eqref{Udvali}.
We can improve on this description, by employing the Quantum Mechanical theory of the 
harmonic oscillator as a (better) mean field approximation for the Newtonian gravitational
interaction acting on each boson inside the BEC.
The potential $U$ in Eq.~\eqref{Udvali} is therefore replaced by~\footnote{This is nothing but
Newton oscillator, which would correspond to a homogenous BEC distribution in the
Newtonian approximation.}
\be
V
&=&
\frac{1}{2}\,\mu\,\omega^2\,(r^2-d^2)\,\Theta(d-r)
\nonumber
\\
&\equiv&
V_0(r)\,\Theta(d-r)
\ ,
\label{Vho}
\ee
and we further set $V(0)=U(0)$, so that
\be
\frac{1}{2}\,\mu \, \omega^2 \, d^2 = N\,\alpha\,\frac{ \hbar}{\ell}
\ .
\ee
We also assume that the effective mass, length and frequency of a single graviton mode
are related by $\mu = \hbar\, \omega = {\hbar}/{\ell}$, which immediately leads to
\be
d
=\sqrt{2\, N\, \alpha} \, \ell
=\sqrt{2\, N} \, \lp
\ .
\label{dN}
\ee
\par
If we neglect the finite size of the well, the Schr\"odinger equation in polar coordinates,
\be
\frac{\hbar^2}{2\,\mu\,r^2}
\left[
\frac{\partial}{\partial r} \left( r^2 \,\frac{\partial}{\partial r} \right)
+
\frac{1}{\sin \theta}\, \frac{\partial}{\partial \theta}
\left(  \sin \theta\, \frac{\partial}{\partial \theta} \right)
+ \frac{1}{\sin^2 \theta} \,\frac{\partial^2}{\partial \phi^2}
\right]
\psi
=
(V_0-E)\,
\psi
\ ,
\label{schrod}
\ee
yields the well-known eigenfunctions
\be
\psi_{nlm}(r,\theta,\phi)
=
\mathcal{N}\,r^l\, e^{-\frac{r^2}{ 2\, \ell^2}}\, _1F_1(-n,l+3/2,r^2/\ell^2)\, Y_{lm}(\theta,\phi)
\ ,
\label{psin}
\ee
where $\mathcal{N}$ is a normalization constant, $_1F_1$ the Kummer confluent
hypergeometric function of the first kind and $Y_{lm}(\theta,\phi)$
are the usual spherical harmonics.
The corresponding energy eigenvalues are given by
\be
E_{nl}
&=&
\hbar\, \omega \left[2\,n+l + \frac{3}{2} - V(0) \right]
\nonumber
\\
&=&
\hbar\,\omega\left[2\,n+l + \frac{1}{2}\left(3-\frac{d^2}{\ell^2}\right)\right]
\ ,
\ee
where $n$ is the radial quantum number and $l$ the angular momentum
(not to be confused with $\ell$).
Following the idea in Ref.~\cite{gomez}, we view the above spectrum as
representing the effective Quantum Mechanical dynamics of depleting modes, which can be described
by the first (non-rotating) excited state~\footnote{Note we have already integrated
out the angular coordinates.}
\be
\psi_{100}(r)
=
\sqrt{\frac{2}{3\,\ell^7\,\sqrt{\pi}}}\,
e^{-\frac{r^2}{ 2\, \ell^2}}\, 
\left(2\,r^2-3\,\ell^2\right)
\ .
\label{psi1}
\ee
The marginally binding condition~\eqref{energy0}, that is $E_{10}\simeq 0$, then
leads to the scaling laws
\be
\ell
=
\sqrt{\frac{2\,N}{7}}\, \lp
\quad
{\rm and}
\quad
\mu
=
{\sqrt{\frac{7}{2\,N}}} \,\mpl
\ ,
\ee
in perfect qualitative agreement with Eq.~\eqref{Max}.
\par
We can now estimate the effect of the finite width of the potential
well~\eqref{Vho} by simply applying first order perturbation theory
and obtain
\be
\Delta E_{10}
&=&
-\int_d^\infty r^2 \,\ud r \, \psi_{100}^2(r) \, V_0(r)
\nonumber
\\
&\simeq&
-\frac{0.1}{\sqrt{N}}\,\mpl
\ .
\ee
This can now be compared, for example, with the ground state energy
$E_{00}=-\sqrt{14/N}\,\mpl\simeq -3.7\,\mpl/\sqrt{N}$.
Since $|\Delta E_{10}|\ll |E_{00}|$, our approximation appears reasonable.
\par
We however remark that the ground state energy in this model has no physical meaning.
Indeed, the Schr\"odinger equation~\eqref{schrod} must be viewed as describing the
effective dynamics of black hole constituents, and the total energy of the ``harmonic black hole'' is still given
by the sum of the individual boson effective masses,
\be
M=
N \,\mu
\simeq
\sqrt{\frac{7\,N}{2}}\,\mpl
\ ,
\label{Mtot}
\ee
in agreement with the ``maximal packing'' of Eq.~\eqref{Max} and the expected
mass spectrum of quantum black holes (see, for example, Refs.~\cite{Bekenstein:1974jk,Dvali:2011nh}).
\section{Regular geometry}
\label{secReg}
\setcounter{equation}{0}
It is now reasonable to assume that the actual density profile of the BEC gravitational source
is related to the ground state wave function in Eq.~\eqref{psin} according to
\be
\rho(r)
\simeq
M\,\psi_{000}^2
\simeq
\frac{7^2\,\mpl\,e^{-\frac{7\,r^2}{2\,N\,\lp^2}}}{\sqrt{\pi}\,N\,\lp^3}
\ .
\label{rhobec}
\ee
Similar Gaussians profiles have been extensively studied in
Refs.~\cite{Nicolini:2005vd, Nicolini},
where it was proven that such densities satisfy the Einstein field equations
with a ``de~Sitter vacuum'' equation of state, $\rho = - p$,
where $p$ is the pressure.
Curiously, BECs can display this particular equation of state~\cite{stringari}.
This feature provides a connection between Quantum Mechanics and the geometrical description.
\par
Let us indeed take the static and normalised, energy density profile of
Ref.~\cite{Nicolini:2005vd},~\footnote{The squared length $\theta$ should not
be confused with one of the angular coordinates of the previous expressions.
Also, note $\rho$ has already been integrated over the angles.}
\be
\rho(r)
=
\frac{M\,e^{-\frac{r^2}{4\,\theta}}}{\sqrt{4\,\pi}\ \theta^{3/2}} \label{rhopiero}
\ ,
\ee
where $\sqrt{\theta}$ is viewed as a fundamental length
related to space-time non-commutativity, and $r$ is the radial coordinate
such that the integral inside a sphere of area $4\,\pi\,r^2$,
\be
M(r)
=
\int_0^r \rho(\bar r)\,\bar r^2\,\ud \bar r
=
M\, \frac{\gamma(3/2,r^2/4\theta)}{\Gamma(3/2)}
\ ,
\label{M}
\ee
gives the total Arnowitt-Deser-Misner (ADM) mass $M$ of the object for $r\to\infty$.
In the above, $\Gamma(3/2)$ and $\gamma(3/2,r^2/4\theta)$ are the complete
and upper incomplete Euler Gamma functions, respectively.
This energy distribution then satisfies Einstein field equations together with the
Schwarzschild-like metric
\be
\ud s^2
=
-f(r)\,\ud t^2
+ f^{-1}(r)\,\ud r^2
+ r^2\, \ud\Omega^2
\ ,
\ee
where
\be
f(r)= 1 - \frac{2\,\gn\, M(r)}{r}
\ .
\ee
According to Ref.~\cite{Nicolini:2005vd}, one has a black hole only if the
\emph{mass-to-characteristic length\/} ratio is sufficiently large,
namely for
\be
M
\gtrsim
1.9\,\frac{\sqrt{\theta}}{\gn}
=
1.9\,\mpl\,\frac{\sqrt{\theta}}{\lp}
\equiv
M_*
\ .
\label{Mth}
\ee
If the above inequality is satisfied, the metric function $f=f(r)$ has two zeros and
there are two distinct horizons.
For $M=M_*$, $f=f(r)$ has only one zero which corresponds to an ``extremal''
black hole, with two coinciding horizons (and vanishing Hawking temperature).
The latter represents the minimum mass black hole, and a candidate
black hole remnant of the Hawking decay~\cite{nc}.
Further, the classical Schwarzschild case is precisely recovered in the limit
$\gn\,M/\sqrt{\theta} \to\infty$, so that departures from the standard geometry
become quickly negligible for very massive black holes.
\par
Going back to the BEC model, whose total ADM mass is given in Eq.~\eqref{Mtot},
and comparing the Gaussian profile~\eqref{rhobec} with Eq.~\eqref{rhopiero},
that is setting $\theta=N\,\lp^2/14$,
one finds that the condition in Eq.~\eqref{Mth} reads
\be
1.8\,\sqrt{N}
\gtrsim
0.5\,\sqrt{N}
\ ,
\ee
and is always satisfied (for $N\ge 1$).
We can therefore conclude that harmonic black holes always have two horizons,
and the degenerate case is not realised in their spectrum.
Although this mismatch might appear as a shortcoming of our construction,
it is actually consistent with the idea that the extremal case should have vanishing
Hawking temperature and therefore no depleting modes.
It also implies that the final evaporation phase, if it ends in the extremal case,
must be realised by a transition that most likely drives the BEC out of the critical point.
The precise nature of such a ``quantum black hole'' state remains, however,
unclear (see, for example, Refs.~\cite{X}).
\section{Conclusions and outlook}
\label{secConc}
\setcounter{equation}{0}
We have shown that the scenario of Ref.~\cite{gomez}, in which black hole inner degrees of freedom
(as well as the Hawking radiation) correspond to depleting states in a BEC, can be understood and
recovered in the context of General Relativity by viewing a black hole as made of the superposition
of $N$ constituents, with a Gaussian density profile, whose characteristic length is given by the
 constituents' effective Compton wavelength.
From the point of view of Quantum Mechanics, such states straightforwardly arise from a binding
harmonic oscillator potential.
Moreover, requiring the existence of (at least) a horizon showed that the extremal case,
corresponding to a remnant with vanishing Hawking temperature, is not realised in the
harmonic spectrum~\eqref{Mtot}.
Such states will therefore have to be described by a different model. 
\par
At the threshold of black hole formation (see, for example, Ref.~\cite{orlandi} and
References therein), for a total ADM mass $M\simeq \mpl$ (thus $N\simeq 1$),
the above description should allows us to describe Quantum Mechanical processes
involving black hole intermediate (or metastable) states.
In order to estimate the typical life-times of such quantum black holes, a better
approximation of the potential outside the characteristic size of the
object will likely be needed~\footnote{For example, one might adapt
the construction yielding the effective potential acting on collapsing
nested shells obtained in Refs.~\cite{alberghi}.}.
However, we can already anticipate that quantum black holes with spin should be
relatively easy to accommodate in our description, by simply considering
states in Eq.~\eqref{psin} with $l>0$.
This should allow us to consider more realistic quantum black hole formation from particle
collisions, since particles most likely scatter with non-zero impact
parameter.
\par
Many questions are still left open.
First of all, the discretisation of the mass has an important consequence in the classical limit.
For example, let us look again at Eq.~\eqref{Mtot}, and consider two non-rotating black holes
with mass $M_1=\sqrt{\frac{7}{2}\,N_1} \, \mpl$ and $M_2=\sqrt{\frac{7}{2}\, N_2} \, \mpl$,
where $N_1$ and $N_2$ are positive integers, which slowly merge in a head-on collision
(with zero impact parameter).
The resulting black hole should have a mass $M$ which is also given by Eq.~\eqref{Mtot}.
However, there is in general no integer $N_3$ such that
$\sqrt{N_3} = \sqrt{N_1} + \sqrt{N_2}$.
It therefore appears that either the mass should not be conserved, $M\not= M_1+M_2$,
or the mass spectrum described by Eq.~\eqref{Mtot} is not complete.  
This problem, which is manifestly more significant for small black hole masses
(or, equivalently, integers $N$), is shared by all those models in which the the black hole
mass does not scale exactly like an integer.
If we wish to keep Eq.~\eqref{Mtot}, or any equivalent mass spectrum, we might then argue
that a suitable amount of energy (of order $M_1+M_2-M_3$)
should be expelled during the merging, in order to accommodate the overall mass
into an allowed part of the spectrum.
In this case, one may also wonder if this emission can be thought of as
some sort of Hawking radiation~\footnote{Note that for vanishing impact parameter,
one does not expect any emission of classical gravitational waves.},
or if it is completely different in nature.
\par
Another issue regards the assumption in Eq.~\eqref{rhobec}, i.e.~the idea that the
classical density profile corresponds to the square modulus of the (normalised)
wavefunction.
At the semiclassical level, this seems reasonable and intuitive, but necessarily
removes the concept of ``point-like test particle'' from General Relativity,
thus forcing us to reconsider the idea of geodesics only in terms of propagation
of extended wave packets,
which might show unexpected features or remove others from the classical theory.
Also, elementary particles would not differ from extended massive objects and therefore
should have an equation of state (see, for instance, the old shell model in
Refs.~\cite{adm60}).
Would this equation of state be an observable and enter the description of the particle
on the same level as any other quantum number?
Do different particles have different equations of state?
\par
Last but not least, there is the question of describing the formation of a
BEC during a stellar collapse.
Condensation is usually achieved at extremely low temperature,
when the thermal de~Broglie wavelength becomes comparable to the
inter-particle spacing.
Whereas one has no doubt that particles inside a black hole are extremely
packed, it is not clear how such a dramatic drop of temperature could occur.
One might find a reason for this in some modification of the laws of thermodynamics
inside the event horizon.
\section*{Acknowledgements}
This work is supported in part by the European Cooperation
in Science and Technology (COST) action MP0905 ``Black Holes in a Violent  Universe". 
\end{document}